\documentstyle[aasms4,psfig]{article}

\newcommand\beq{\begin{equation}} 
\newcommand\eeq{\end{equation}}

\received{} 
\accepted{} 

\lefthead{Ma et al.} 
\righthead{SED, Age and Metallicity of Star Clusters in M33} 

\begin{document}

{ \title{SED, Age and Metallicity of Star Clusters in M33}

\author{ 
Jun Ma\altaffilmark{1}, 
Xu Zhou\altaffilmark{1},
Xu Kong\altaffilmark{2},
Hong Wu\altaffilmark{1},
Jiansheng Chen\altaffilmark{1}, 
Zhaoji Jiang\altaffilmark{1}, 
Jin Zhu\altaffilmark{1},
and
Suijian Xue\altaffilmark{1}
}

\altaffiltext{1}{National Astronomical Observatories, 
Chinese Academy of Sciences, Beijing, 100012, P. R. China;
majun@vega.bac.pku.edu.cn}

\altaffiltext{2}{Center for Astrophysics, University of Science and
        Technology of China, Hefei, 230026, P. R. China}
\authoremail{majun@vega.bac.pku.edu.cn}

\begin{abstract}

In this paper we present CCD
spectrophotometry of the star
clusters that were detected by Chandar, Bianchi, \&
Ford in the nearby
spiral galaxy M33, using the images obtained with the Beijing Astronomical
Observatory $60/90$ cm Schmidt Telescope in 13 intermediate-band filters
from 3800 to 10000{\AA}.  The observations
cover the whole area of M33
with a total integration of 32.75 hours from September
23, 1995 to August 28, 2000.
This provides a multi-color map of M33 in pixels
of $1\arcsec{\mbox{}\hspace{-0.15cm}.} 7
\times 1\arcsec{\mbox{}\hspace{-0.15cm}.} 7$.
By aperture photometry, we obtain the
spetral energy distributions (SEDs)
of these star clusters. 
Using theoretical stellar population synthesis
models, we also obtain the distributions of
age and metallicity of these star clusters. These
clusters formed continuously from $\sim 3\times10^6$ -- $10^{10}$ years,
and have a large
span of metallicity from ${\rm {Z}=0.0004}$ to ${\rm {Z}=0.05}$.

\end{abstract}

\keywords{galaxies: individual (M33) -- galaxies: evolution -- galaxies:
star clusters}

\section{INTRODUCTION}

The importance of the study of star clusters is difficult to
overstate, especially in Local Group galaxies. Star
clusters, which represent, in distinct and luminous ``packets'',
single age and single abundance points, and encapsulate at least
a partial history of the parent galaxy's evolution,
can provide a unique laboratory for studying
the ongoing and past star formation in the parent galaxy.
Determination of elemental abundances in cluster stars
is crucial to our understanding of the chemical evolution
and star formation histories of the galaxy.
In addition, studies of star cluster populations
could help our understanding of the relationships between
cluster formation and the physical morphology of the parent
galaxy.

M33 is a small Scd Local Group galaxy, about 15 times farther from
us than the LMC (distance modulus is 24.64) (Freedman,
Wilson, \& Madore 1991; Chandar, Bianchi, \&
Ford 1999a). It is interesting and important because
it represents a morphological type intermediate
between the largest ``early-type'' spirals and the dwarf
irregulars in the Local Group (Chandar, Bianchi, \& Ford 1999a). 
Our collaboration, the Beijing-Arizona-Taiwan-Connecticut
(BATC) Multicolor Sky Survey (Fan et al. 1997; Zheng et al. 1999),
already had this spiral galaxy as part of its galaxy
calibration program. The BATC program uses the 60/90 cm
Schmidt telescope at the Xinglong Station of Beijing Astronomical
Observatory, with its focal plane equipped with a
$2048 \times 2048$ Ford CCD, and has custom designed a set of
15 intermediate-band filters to do spectrophotometry for
preselected 1 deg$^{2}$ regions of the northern sky with this
CCD system.

For M33, a database of star clusters have been
yielded from the ground-based work (Hiltner 1960;
Kron \& Mayall 1960; Christian \& Schommer 1982, 1988;
Melnick \& D'Odorico 1978), and from the {\it {Hubble Space Telescope
(HST)}} images (Chandar, Bianchi, \&
Ford 1999a). Especially, Chandar, Bianchi, \& Ford (1999a) presented the
first unbiased, representive sample of star clusters, sampling
a variety of environments from outer regions to spiral
arms and central regions, and can be used to probe the global
properties of M33.

Since the pioneering work of Tinsley (1972) and Searle et al. (1973),
evolutionary population synthesis has become a standard technique to
study the stellar populations of galaxies. This is a result of the improvement
in the theory of the chemical evolution of galaxies, star formation,
stellar evolution and atmospheres, and of the development of synthesis
algorithms and the availability of various evolutionary synthesis models.
A comprehensive compilation of such models was presented by Leitherer et
al. (1996) and Kennicutt (1998). More widely used models are those from
the Padova and Geneva group (e.g. \cite{Schaerer97}; \cite{Schaerer98};
\cite{Bressan96}; \cite{Chiosi98}), GISSEL96 (\cite{Charlot91};
\cite{Bruzual93}; \cite{Bruzual96}), PEGASE (\cite{Fioc97}) and
STARBURST99 (\cite{Leitherer99}).

In this paper, we present the SEDs of
the star clusters that were detected by Chandar, Bianchi, \&
Ford (1999a) in M33,
and study the distributions of age and metallicity of these clusters
by using the theoretical evolutionary population synthesis methods.
The multi-color photometry is powerful
to provide  the accurate SEDs for these
stellar clusters.

The outline of the paper is as follows.  Details of observations
and data reduction are given in section 2. In section 3, we provide
a brief description of the 
stellar population synthesis models of Bruzual \&
Charlot (1996). The distributions of metallicity and age
are given in section 4.
The summary and discussion are presented in section 5.

\section{SAMPLE OF STAR CLASTERS, OBSERVATIONS AND DATA REDUCTION}

\subsection{Sample of Star Clusters}

The sample of star clusters in this paper is from
Chandar, Bianchi, \& Ford (1999a), who used 20 multiband
$Hubble$ $Space$ $Telescope$ $(HST)$ WFPC2 fields to
search for star clusters much closer to the nucleus of M33
than previous studies. Since these star clusters
populate the variety of environments from the  outer regions to spiral
arms and central regions, they can be used to probe the global
properties of the parent galaxy. At the same time, the accurate positions
are presented in Table 2 of Chandar, Bianchi, \& Ford (1999a).
So, as a first step, we select these star clusters to be studied,
and obtain their SEDs by aperture photometry. The
distribution of age and metallicity for these star clusters
are derived by using the theoretical evolutionary
population synthesis methods. Clusters 17, 39, 41 and 42
are not included in our sample
because of their low signal-to-noise ratio in the images
of some BATC filters.

\subsection{CCD Image Observation}

The large field multi-color observations of the spiral galaxy M33 were
obtained in the BATC photometric system. The telescope used is the
60/90 cm f/3 Schmidt Telescope of Beijing Astronomical Observatory (BAO),
located at the Xinglong station. A Ford Aerospace 2048$\times$2048 CCD
camera with 15 $\mu$m pixel size is mounted at the Schmidt focus of the
telescope. The field of view of the CCD is $58^{\prime}$ $\times $ $
58^{\prime}$ with a pixel scale of $1\arcsec{\mbox{}\hspace{-0.15cm}.} 7$.  

The multi-color BATC filter system includes 15 intermediate-band filters,
covering the total optical wavelength range from 3000 to 10000{\AA}
(see Fan et al. 1996). The filters were specifically designed to avoid
contamination from the brightest and most variable night sky emission
lines. A full description of the BAO Schmidt telescope, CCD, data-taking
system, and definition of the BATC filter systems are detailed elsewhere
(\cite{Fan96}; \cite{Zheng99}).
The images of M33 covering
the whole optical body of M33 were accumulated in 13 intermediate band
filters with a total exposure time of about 32.75 hours from September
23, 1995 to August 28, 2000.  The CCD images are centered at ${\rm
RA=01^h33^m50^s{\mbox{}\hspace{-0.13cm}.}58}$ and
DEC=30$^\circ39^{\prime}08^{\prime\prime}{\mbox{}\hspace{-0.15cm}.4}$
(J2000). The dome flat-field images were taken by using a diffuse plate in
front of the correcting plate of the Schmidt telescope. For flux calibration,
the Oke-Gunn primary flux standard stars HD19445, HD84937, BD+262606
and BD+174708 were observed during photometric nights. The parameters of
the filters and the statistics of the observations are given in Table 1.
Figure 1 is ``True-color" estimate of M33 generated by using
the BATC03 (4210{\AA}) filter image for blue, BATC07 (5785{\AA}) for
green, and BATC10 (7010{\AA}) for red.

\begin{figure}
\figurenum{1}
\vspace{0.5cm}
\caption{``True-color" estimate of M33 generated by using
the BATC03 (4210{\AA}) filter image for blue, BATC07 (5785{\AA}) for
green, and BATC10 (7010{\AA}) for red; the image is balanced by making
the background old population orange and hot stars blue.  The center
(origin) of the image is located at
${\rm RA=01^h33^m50^s{\mbox{}\hspace{-0.13cm}.}58}$
DEC=30$^\circ39^{\prime}08^{\prime\prime}{\mbox{}\hspace{-0.15cm}.4}$
(J2000.0). North is up and east is to the left.
}
\label{fig1}
\end{figure}

\subsection{ Image Data Reduction }

The data were reduced with standard procedures, including bias subtraction
and flat-fielding of the CCD images, with an automatic data reduction
software named PIPELINE 1 developed for the BATC multi-color sky survey
(\cite{Fan96}; \cite{Zheng99}).
The flat-fielded images of each color were combined
by integer pixel shifting. The cosmic rays and bad pixels were corrected
by comparison of multiple images during combination. The images were
re-centered and position-calibrated using the $HST$ Guide Star Catalog.
The absolute flux of intermediate-band filter images was
calibrated using observations of standard stars. Fluxes as observed
through the BATC filters for the Oke-Gunn stars were derived by convolving
the SEDs of these stars with the measured BATC filter transmission
functions (\cite{Fan96}). {\it Column} 6 in Table 1 gives the zero point
error, in magnitude, for the standard stars in each filter. The formal
errors we obtain for these stars in the 13 BATC filters are $\la 0.02$
mag. This indicates that we can define the standard BATC system to an
accuracy of  $\la 0.02$ mag.

\subsection{Integrated Photometry}

For each star cluster, aperture photometry was used to obtain
magnitudes. For avoiding contamination from nearby objects, a smaller aperture
of $6\arcsec{\mbox{}\hspace{-0.15cm}.} 8$,
which corresponds
to a diameter of 4 pixels in Ford CCDs, was adopted.
Considering the large seeing of the Xinglong station, aperture
corrections were computed using isolated stars.
Since these star clusters sample a variety of environments from
outer regions to spiral arms and central regions, background
subtraction is difficult. We determined the background in annulus
of from 2 to 5 pixels from each star cluster.
In this annulus, we fitted each row of the
image by linear  median fitting to obtain a surface.
Then, we repeat this process in the
column direction of the surface that is obtained in the row of fitting.
After this fit, we reject points
higher and lower than 30 percent of the mean background.
Finally, we obtained the final smoothed surface of background.
Using this surface of background, we made the
background subtraction for each cluster.
The spectral energy
distributions (SEDs) for the 56 star clusters
were obtained.
Table 2 contains the following information: {\it Column 1} is cluster
number which is taken from Chandar, Bianchi, \& Ford (1999a).
{\it Column} 2 to {\it Column} 14 show the magnitudes of
different bands. Second line of each star cluster is
the uncertainties of magnitude of corresponding band.
The uncertainties for each filter take into account the
error from the object count rate, sky variance, and instrument
gain.

\subsection{Comparison with Previous Photometry}

Using the Landolt standards, Zhou et al. (2001) presented the relationships
between the BATC intermediate-band system and UBVRI broadband system
by the catalogs of Landolt (1983, 1992) and Galad\'\i-Enr\'\i quez et al. (2000).
We show the coefficients of the fits
in equations 1 and 2.
\beq
m_B=m_{04}+(0.2218\pm0.033)\times(m_{03}-m_{05})+0.0741\pm0.033,
\eeq
\beq
m_V=m_{07}+(0.3233\pm0.019)\times(m_{06}-m_{08})+0.0590\pm0.010.
\eeq
Using equations 1 and 2, we transformed the magnitudes of
the star clusters in BATC03, BATC04 and BATC05 bands to ones in B band,
and in BATC06, BATC07 and BATC08 bands to ones in V band.
For clusters 1, 7, 37, 40 and 43,
we change $m_{05}$ to $m_{04}$ because of the strong emission
or absorption lines in BATC05 band.
Figure 2 plots the comparison of
$V$ (BATC) and ($B$$-$$V$) (BATC) photometry with previously
published measurements
(Chandar, Bianchi, \& Ford 1999a). Table 3 also shows this
comparison. The mean $V$ magnitude and color differences (this paper
$-$ the paper (of Chandar, Bianchi, \& Ford 1999a)) are $<\Delta V>
=-0.009\pm0.025$ and
$<\Delta (B-V)>= -0.179\pm0.039$, respectively.
The uncertainties in $B$ (BATC) and $V$ (BATC) have been added linearly,
i.e. $\sigma_B=\sigma_{04}+0.2218\times(\sigma_{03}+\sigma_{05})$, and
$\sigma_B=\sigma_{07}+0.3233\times(\sigma_{06}+\sigma_{08})$, to reflect the
error in the three filter measurements.
For the colors, we also added the errors linearly,
i.e. $\sigma_{(B-V)}=\sigma_B+\sigma_V$. 
{\begin{figure}
\figurenum{2}
\centerline{\psfig{file=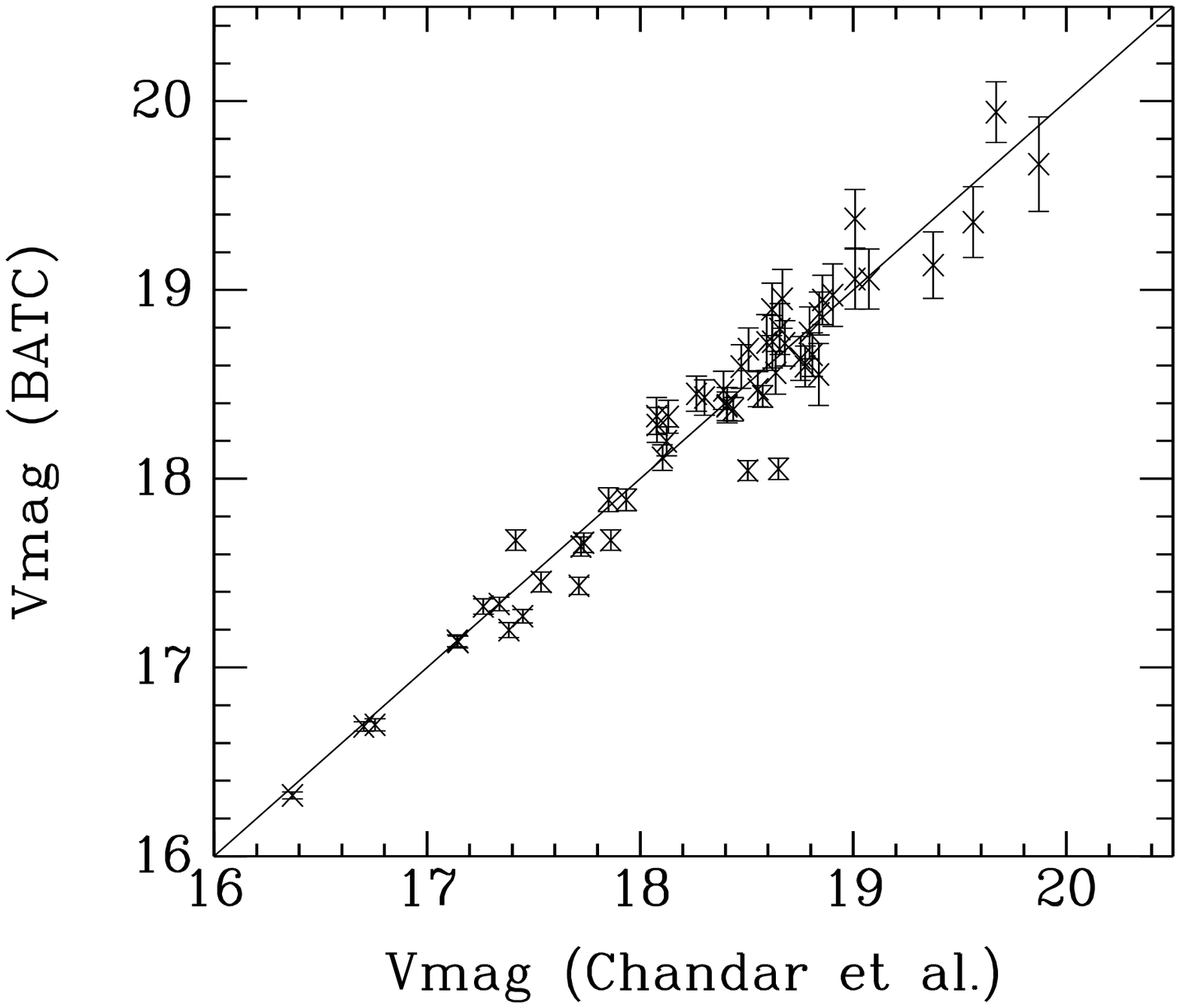,width=16.0cm,angle=-90}}
\label{fig2}
\end{figure}
\begin{figure}
\vspace{-2cm}
\figurenum{2}
\centerline{\psfig{file=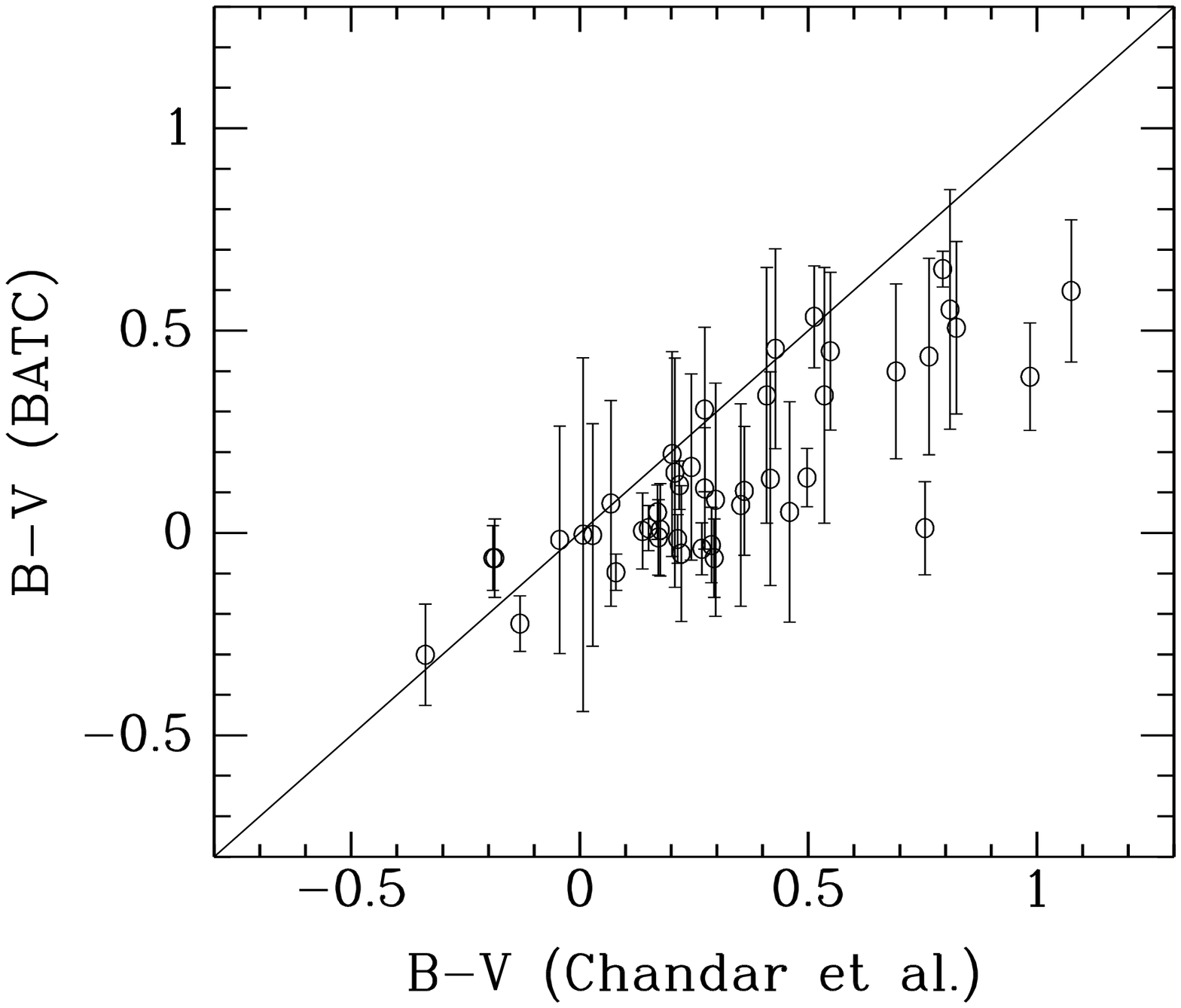,width=16.0cm,angle=-90}}
\vspace{-1cm}
\caption{Comparison of Cluster Photometry with Previous Measurements}
\label{fig2}
\end{figure}}

\section{DATABASES OF SIMPLE STELLAR POPULATIONS}

A simple stellar population (SSP) is defined as a single generation
of coeval stars with fixed parameters such as metallicity, initial
mass function, etc. (\cite{Buzzoni97}).
SSPs are the basic building blocks of synthetic spectra
of galaxies that can be used to infer the formation and subsequent
evolution of the parent galaxies (\cite{Jab96}).
They are modeled by a collection of stellar evolutionary tracks with
different masses and initial chemical compositions, supplemented
with a library of stellar spectra for stars at different evolutionary
stages in evolution synthesis models.
In order to study the
integrated properties of star clusters in M33, as the first step,
we use the SSPs of Galaxy Isochrone Synthesis Spectra Evolution Library
(Bruzual \& Charlot 1996 hereafter GSSP) because
they are simple and reasonably well understood.
 
\subsection{Spectral Energy Distribution of GSSPs}

The Bruzual \& Charlot (1996) study has extended the Bruzual \& Charlot
(1993) evolutionary population synthesis models. The updated version
provides the evolution of the spectrophotometric properties for a wide
range of stellar metallicity. They are based on the stellar evolution
tracks computed by Bressan et al. (1993), Fagotto et al. (1994), and
by Girardi et al. (1996), who use the radiative opacities of Iglesias
et al. (1992). This library includes tracks for stars with metallicities
$Z=0.0004, 0.004, 0.008, 0.02, 0.05,$ and $0.1$, with the helium abundance
given by $Y=2.5Z+0.23$ (The reference solar metallicity is $Z_\odot=0.02$.).
The stellar spectra library is from Lejeune et al. (1997, 1998) for all
the metallicities listed above, which in turn consists of Kurucz (1995)
spectra for the hotter stars (O-K), Bessell et al. (1991) and Fluks
et al. (1994) spectra for M giants, and Allard \& Hauschildt (1995)
spectra for M dwarfs. GSSP models assume an
initial Salpeter (1955) mass function $\Phi(M)=A \times M^{-\alpha}$
with $\alpha=2.35$ and normalization constant
$A=1$, and a lower cutoff
$M_{\rm l}=0.1M_{\odot}$ and an upper cutoff $M_{\rm u}=125M_{\odot}$
(\cite{Sawicki98}).

\subsection{Integrated Colors of GSSPs}

Kong et al. (2000) have obtained the age, metallicity, and interstellar-medium
reddening distribution for M81.
They found the best match between the intrinsic colors
and the predictions of GSSP for each cell of M81.
To determine the distributions of age and metallicity of the
star clusters in M33, we follow the method of Kong et al. (2000).
Since the observational
data are integrated luminosity, we need to convolve
the SED of GSSP with BATC filter profiles to obtain the optical
and near-infrared integrated luminosity for comparisons (Kong et al. 2000).
The integrated luminosity
$L_{\lambda_i}(t,Z)$ of the $i$th BATC filter can be calculated with
\beq 
L_{\lambda_i}(t,Z) =\frac{\int
F_{\lambda}(t,Z)\varphi_i(\lambda)d\lambda} {\int
\varphi_i(\lambda)d\lambda},
\eeq
where $F_{\lambda}(t,Z)$ is the spectral energy distribution of
the GSSP of metallicity $Z$ at age $t$, $\varphi_i(\lambda)$ is the
response functions of the $i$th filter of the BATC filter system
($i=3, 4, \cdot\cdot\cdot, 15$),
respectively.

The absolute luminosity can be obtained if we know the distance to M33
and the extinction along the line of sight.
However, we do not know these parameters exactly. So, we should
work with the colors because of their indenpance of the distance.
We calculate the integrated colors of a GSSP relative to
the BATC filter BATC08 ($\lambda=6075${\AA}):

\beq 
\label{color}
C_{\lambda_i}(t,Z)={L_{\lambda_i}(t,Z)}/{L_{6075}(t,Z)}.  
\eeq

As a result, we obtain intermediate-band colors for 6 metallicities from
$Z=0.0004$ to $Z=0.1$.

\section{REDDENING CORRECTION AND DISTRIBUTIONS OF METALLICITY, AGE}

In general, the SED of a stellar system depends on age, metallicity
and reddening along the line of sight. The effects of age, metallicity
and reddening are difficult to separate (e.g. \cite{Calzetti97};
\cite{Origlia99}; \cite{Vazdekis97}). Older age, higher metallicity or
larger reddening all lead to redder SEDs of stellar systems in the
optical (\cite{Molla97}; \cite{Bressan96}). In order to obtain
intrinsic colors for these star clusters, we must correct for
reddening.

\subsection{Reddening Correction}

In order to obtain intrinsic colors of 56 clusters and hence
accurate ages,
the photometric measurements must be dereddened.
The observed colors are affected by two sources of reddening: (1)
the foreground extinction in our Milky Way and (2) internal
reddening due to varying optical paths through the disk of the
parent galaxy. McClure \& Racine (1969) has measured the foreground
color excess, $0.03\pm0.02$ for M33. As Chandar, Bianchi, \&
Ford (1999b) did, we also adopted this value.
For internal reddening of the star clusters, we adopted the values
in the third column of Table 3 of Chandar, Bianchi, \&
Ford (1999b). Besides,
we adopted the extinction curve presented by Zombeck (1990).
An extinction correction $A_{\lambda}=R_{\lambda}E(B-V)$ was
applied, here $R_{\lambda}$ is obtained by interpolating
using the data of Zombeck (1990).

\subsection{Age and Metallicity Distribution}

Since we model the stellar populations of the star clusters
by SSPs, the intrinsic colors for
each star cluster are determined by two parameters: age, and metallicity.
In this section, we will determine these two parameters for these
star clusters simultaneously by a least square method.
The best fit age and metallicity
are found by minimizing the difference between the intrinsic and
integrated colors of GSSP:

\beq 
R^2(n,t,Z)=\sum_{i=3}^{15}[C_{\lambda_i}^{\rm
intr}(n)-C_{\lambda_i}^ {\rm ssp}(t, Z)]^2, 
\eeq
where $C_{\lambda_i}^{\rm ssp}(t, Z)$ represents the integrated color in
the $i$th filter of a SSP with age $t$ and metallicity $Z$,
and $C_{\lambda_i}^{\rm intr}(n)$ is the intrinsic
integrated color for nth star cluster.
From Chandar, Bianchi, \&
Ford (1999b), the distribution of metallicity
of these star clusters is from $\sim 0.0002$ to $0.03$. So,
we only select the modles of three metallicities, 0.0004, 0.004 and 0.02 of
GSSP.

Figure 3 shows the map of the best fit of the integrated color
of a SSP with the intrinsic integrated color for 56 star clusters,
and Table 4 presents the ages and metallicities of these 56 star clusters.
In Figure 3, the thick line represents the integrated
color of a SSP of GSSP, and filled circle represents the intrinsic
integrated color of a star cluster.
From this figure, we also see that clusters 1, 7, 8, 40 and 43
have strong emission lines.
For cluster 2, since the fitting is not good in the above models of the
three metallicities of GSSP (showing by dashed line in Figure 3),
we again select a model of a higher metallicity, 0.05 (showing by the thick line
in Figure 3). We find that this model of metallicity of 0.05 can be
fitted much better for cluster 2.

\begin{figure}
\figurenum{3}
\centerline{\psfig{file=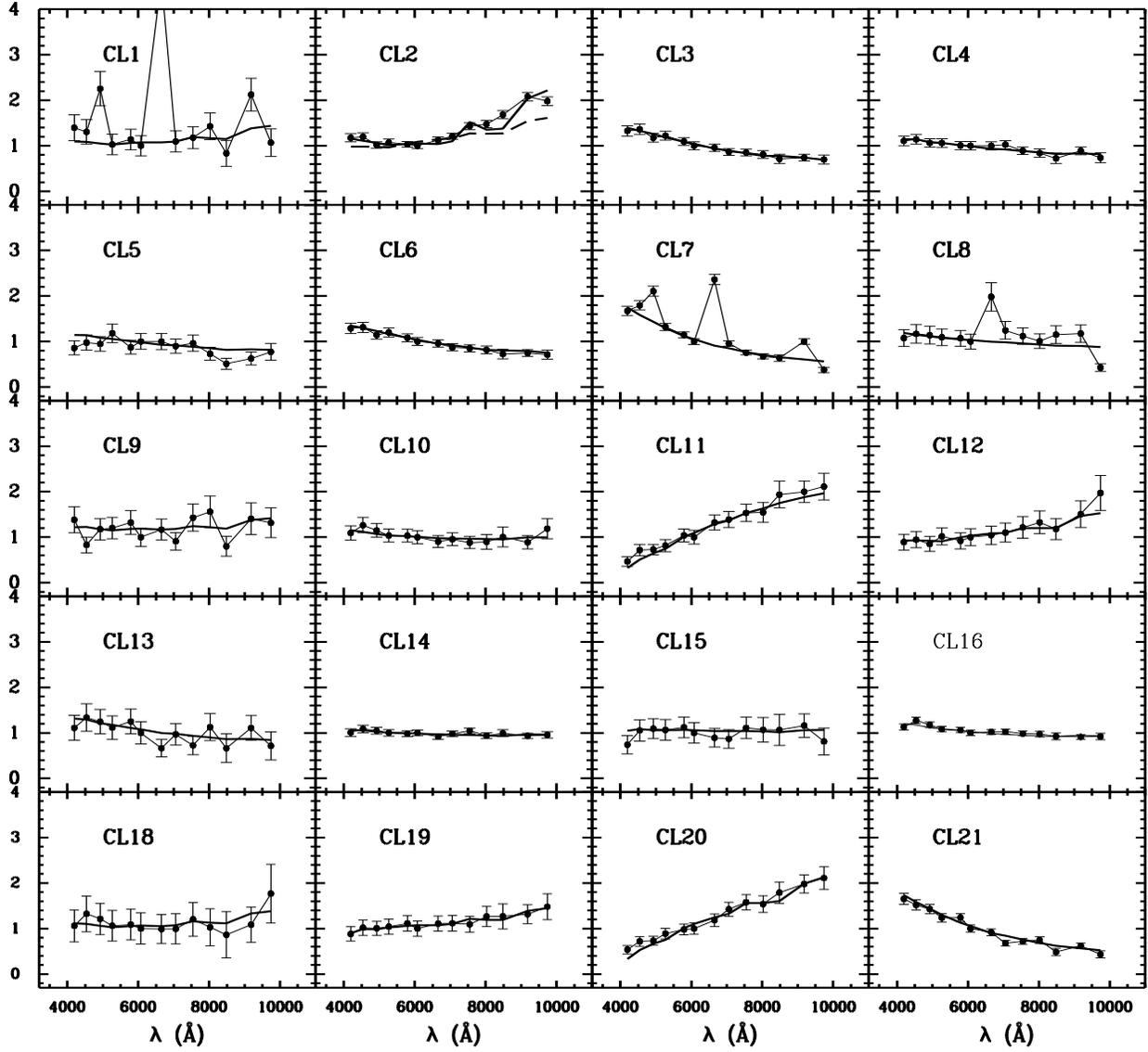,width=22.0cm,angle=270}}
\caption{Map of the best fit of the integrated color
of a SSP with intrinsic integrated color for 56 star clusters.
Thick line represents the integrated color of a SSP, and
filled circle represents the intrinsic integrated color of a star cluster.}
\end{figure}
 
\begin{figure}
\figurenum{3}
\centerline{\psfig{file=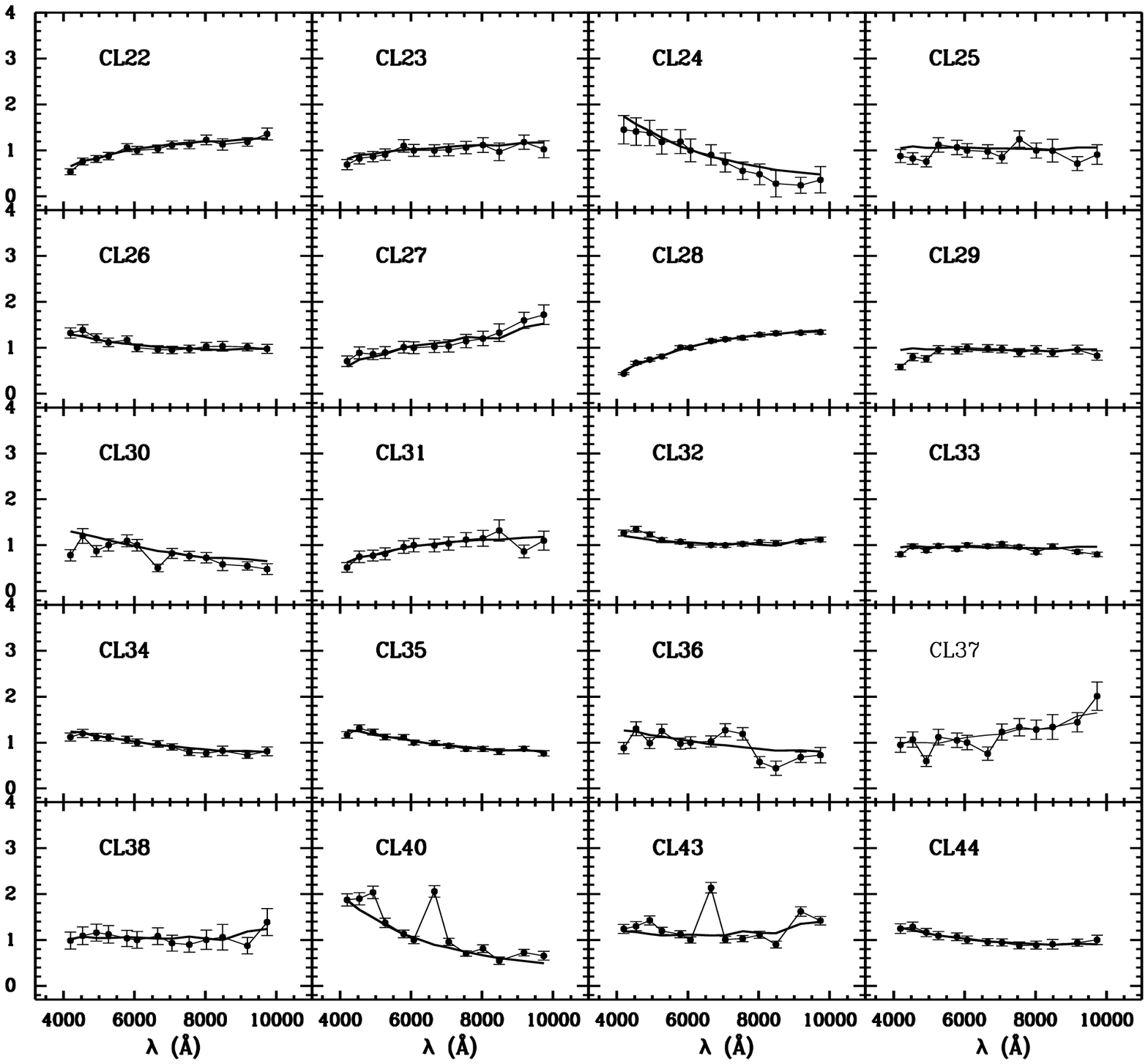,width=22.0cm,angle=270}}
\caption{Continued}
\end{figure}
 
\begin{figure}
\figurenum{3}
\centerline{\psfig{file=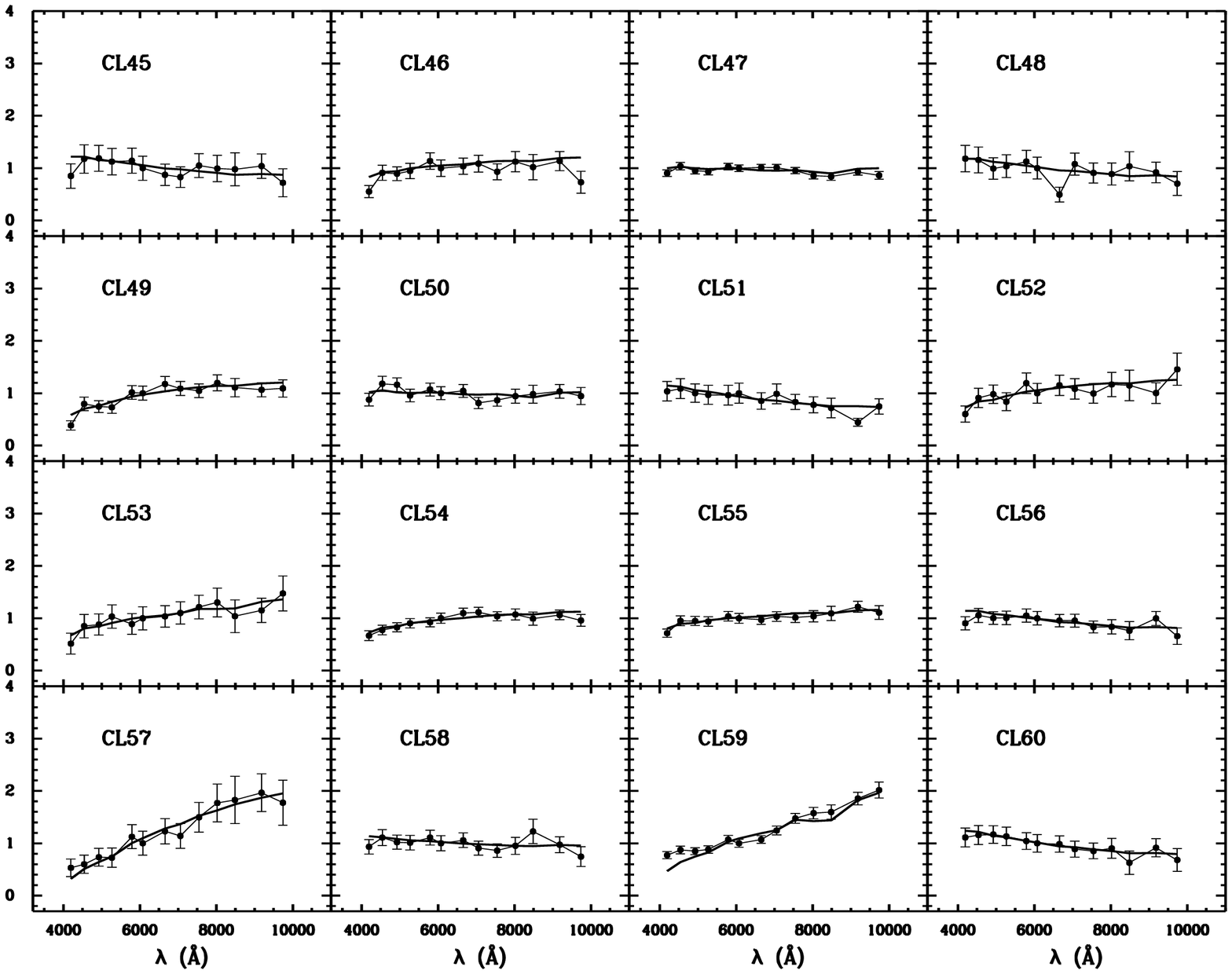,width=20.0cm,angle=270}}
\vspace{-1.0cm}
\caption{Continued}
\end{figure}

Figure 4 presents a histogram of cluster ages. The results show that,
in general, M33 clusters have been forming continuously, with
ages of $\sim 3 \times 10^{6}$ -- $10^{10}$ years. This conclusion
is consistent  with that found by Chandar, Bianchi, \&
Fort (1999b).
There exist three groups of clusters that formed with three
metalicities, $Z=0.02, 0.004$, and $0.0004$. With different metallicities,
the distribution of cluster ages is a little different, too.
With $Z=0.02$ metallicity, the ages of clusters
are younger than $\sim 4 \times 10^{9}$ years.
With $Z=0.004$,
the clusters formed from $\sim 3 \times 10^{6}$ -- $10^{10}$ years.
With $Z=0.0004$, the clusters formed from
$\sim 10^{8}$ -- $10^{10}$ years except of cluster 8.
In this model, there are 15 young clusters ($< 10^{9}$ years). Except of
cluster 26, which is near the center of the host galaxy ($\sim
1'{\mbox{}\hspace{-0.10cm}.} 7$ far from the center), the other
clusters are farther than $4'{\mbox{}\hspace{-0.10cm}.} 0$ from the center
(nearer than $12'{\mbox{}\hspace{-0.10cm}.} 6$). This population of young,
metal-poor clusters appears to not be in the outskirts of the parent
galaxy.
Clusters 11 and 57 have derived ages consistent with that of the globular
clusters of the Milky Way, $\sim 1.5\times 10^{10}$ years. This result is
also consistent with
that found by Chandar, Bianchi, \&Fort (1999b), who
presented clusters 28 and 29 to be as old as $\sim 1.5\times 10^{10}$ years.
From our small sample clusters, we cannot see any evidence
for an adundance gradient. The main reason may be that,
integrated colors of star clusters depend mostly on age, with
a secondary dependence on chemical composition.
So, we can estimate ages of clusters, but cannot determine
metallicities of clusters exactly. As Chandar, Bianchi, \& Ford (1999b, 1999c)
did, we also estimated the ages of our sample clusters by comparing
the photometry of each object with models for different values of metallicity.
Although we presented the metallity of each cluster in Table 4,
we only mean that, in this model of metallicity, the intrinsic integrated color
of each cluster
can do the best fit with the integrated color of a SSP.
\begin{figure}
\figurenum{4}
\centerline{\psfig{file=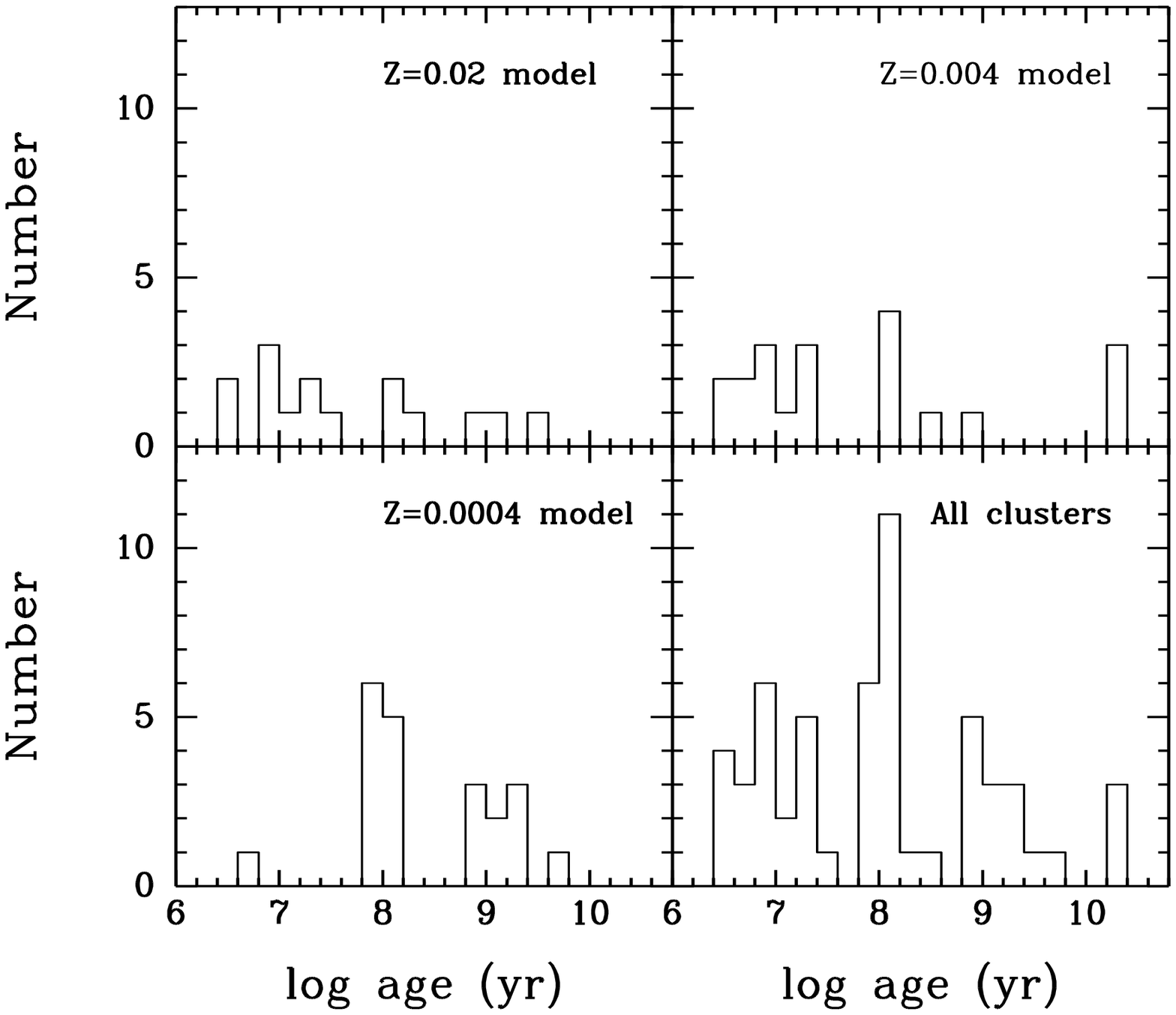,width=24.0cm,angle=-90}}
\vspace{-2.0cm}
\caption{Histogram of M33 cluster ages}
\label{fig4}
\end{figure}

Kong et al. (2000) found a relation between flux ratio of
$I_{8510}\equiv L_{8510}/L_{9170}$ and metallicity for stellar populations
older than 1 Gyr. Using this relation, we can provide an independent
estimate of metallicity for our older clusters without an
assumption of reddening. But, this method can be only used when the
signal-to-noise ratio is high enough. In our sample clusters of M33, the
signal-to-noise ratios are not high enough because of the strong background (These
clusters distribute mainly in the spiral arms and central regions.).
So, the metallicities of the old clusters derived via the technique
of Kong et al. (2000) are also somewhat uncertain.
In our paper, there are 10 clusters, the ages of which are 1 Gyr or older
than 1 Gyr. We calculated their metallicities using the method of Kong et al.
(2000) except of clusters 31, 52 and 57, since the magnitude uncertainties of
these three clusters are too large. Then we obtained their reddenings and ages.
The results are listed in Table 5.
From this table and Table 4, we can see that the metallicities of
clusters 11 and 54 are not consistent by two methods.
We also see that, except of clusters 20 and 59, the cluster ages
derived by two methods, are consistent.
The values of reddening obtained by the method of Kong et al. (2000),
are not consistent with ones of Chandar, Bianchi, \&Fort (1999b) except of
clusters 20, 22 and 28. But, as we know the reddening is difficult
to estimate.

\section{SUMMARY AND DISCUSSION}

In this paper, we have, for the first time, obtained the SEDs
of 56 star clusters of M33 in 13 intermediate colors with
the BAO 60/90 cm Schmidt telescope.
Below, we summarize our main conclusions.

1. Using the images obtained with the Beijing Astronomical
Observatory $60/90$ cm Schmidt Telescope in 13 intermediate-band filters
from 3800 to 10000{\AA}, we
obtained the spectral energy
distributions (SEDs) of the star clusters
that were detected by Chandar, Bianchi, \&
Ford (1999a).

2. Using theoretical stellar population synthesis
models, we obtained the distributions of
age and metallicity of these star clusters. These
clusters formed continuously from $\sim 3\times10^6$ -- $10^{10}$ years,
and have a large
span of metallicity from ${\rm {Z}=0.0004}$ to ${\rm {Z}=0.05}$.

With $\sim 10$ \AA~ resolution spectrophotometry
($\lambda\lambda 3700-5500$ \AA), Christian \& Schommer (1983)
presented that M33 star clusters have a range of ages ($\sim
10^7$ -- $10^{10}$ years). Using the integrated
$UBV$ photometry and IUE $\lambda\lambda1200-3000$ \AA~ spectra, Ciani,
D'Odorico, \& Benvenuti (1984) studied the minuscule ``bulge''
population of M33 and found that, a multigeneration model,
where a young component (age $\sim 10^7$ years) and an old,
metal-poor one (age $\sim 5\times10^9$ years) are superposed,
gives the best fit to the observed data. Schmidt, Bica, \&
Alloin (1990) applied a population synthesis method which uses a
star cluster spectral library and a grid of the star
cluster spectral properties as a function of age and metallicity
(Bica \& Alloin 1986a, b; 1987),
to the blueish nucleus of M33, and gave an age of less than
$5\times10^8$ years for the dominant blue bulge population.
From the histogram of ages in this paper, we can see that
some old clusters in our sample appear to be coeval with the
old population of the bulge.
Chandar, Bianchi, \& Ford (1999b, 1999c) also
estimated ages for the star clusters of this paper by
comparing the $UBV$ and far-UV photometric measurements
(Chandar, Bianchi, \& Ford 1999a, 1999c) to integrated colors
from theoretical models by Bertelli et al. (1994), and found that
these clusters formed continuously from
$\sim 4\times10^6$ -- $10^{10}$ years. Our results are consistent
with the conclusions of Chandar, Bianchi, \& Ford (1999b, 1999c).

To disentangle age, metallicity and reddening in SSPs, we adopted
the $E(B-V)$ values presented in  Chandar, Bianchi, \& Ford (1999b).
These values may be somewhat uncertain. In order to see clearly
how the uncertainties of $E(B-V)$ affect the derived age and
metallicity, we adopt an uncertainty of 0.03 magnitudes of
the $E(B-V)$ values given in  Chandar, Bianchi, \& Ford (1999b).
The results appear that an uncertainty of 0.03 magnitudes in
$E(B-V)$ hardly affects the derived ages and metallicities
on the average, but have a somewhat larger affect on the
older clusters.

\acknowledgments
We would like to thank the anonymous referee for his/her
insightful comments and suggestions that improved this paper.
We are grateful to the Padova group for providing us with a set of
theoretical isochrones and SSPs. We also thank G. Bruzual and
S. Charlot for sending us their latest calculations of SSPs and
for explanations of their code. The BATC Survey is supported by the
Chinese Academy of Sciences, the Chinese National Natural Science
Foundation and the Chinese State Committee of Sciences and
Technology.
The project is also supported in part
by the National Science Foundation (grant INT 93-01805) and
by Arizona State University, the University of Arizona and Western
Connecticut State University.


\begin{table}[ht]
\caption[]{Parameters of the BATC filters and statistics of observations}
\vspace {0.5cm}
\begin{tabular}{cccccc}
\hline
\hline
 No. & Name& cw\tablenotemark{a}~~(\AA)& Exp. (hr)&  N.img\tablenotemark{b}
 & rms\tablenotemark{c} \\
\hline
1  & BATC03& 4210   & 00:55& 04 &0.024\\
2  & BATC04& 4546   & 01:05& 04 &0.023\\
3  & BATC05& 4872   & 03:55& 19 &0.017\\
4  & BATC06& 5250   & 03:19& 15 &0.006\\
5  & BATC07& 5785   & 04:38& 17 &0.011\\
6  & BATC08& 6075   & 01:26& 08 &0.016\\
7  & BATC09& 6710   & 01:09& 08 &0.006\\
8  & BATC10& 7010   & 01:41& 08 &0.005\\
9  & BATC11& 7530   & 02:07& 10 &0.017\\
10 & BATC12& 8000   & 03:00& 11 &0.003\\
11 & BATC13& 8510   & 03:15& 11 &0.005\\
12 & BATC14& 9170   & 01:15& 05 &0.011\\
13 & BATC15& 9720   & 05:00& 26 &0.009\\
\hline
\end{tabular}\\
\tablenotetext{a}{Central wavelength for each BATC filter}
\tablenotetext{b}{Image numbers for each BATC filter}
\tablenotetext{c}{Zero point error, in magnitude, for each filter
as obtained from the standard stars}
\end{table}


{\small
\setcounter{table}{1}
\begin{table}[ht]
\caption{SEDs of 56 Star Clusters}
\vspace {0.3cm}
\begin{tabular}{cccccccccccccc}
\hline
\hline
 No. & 03  &  04 &  05 &  06 &  07 &  08 &  09 &  10 &  11 &  12 &  13 &  14 &  15\\
(1)    & (2) & (3) & (4) & (5) & (6) & (7) & (8) & (9) & (10) & (11) & (12) & (13) & (14)\\
\hline
     1.. & 18.617 & 18.650 & 18.017 & 18.819 & 18.647 & 18.769 & 17.066 & 18.620 & 18.520 & 18.282 & 18.835 & 17.801 & 18.528\\
 & 0.073 &  0.078 &  0.039 &  0.100 &  0.081 &  0.107 &  0.024 &  0.098 &  0.095 &  0.098 &  0.245 &  0.060 &  0.185        \\
     2.. & 16.684 & 16.623 & 16.760 & 16.653 & 16.635 & 16.648 & 16.492 & 16.392 & 16.189 & 16.125 & 15.954 & 15.702 & 15.736\\
 & 0.019 &  0.019 &  0.019 &  0.021 &  0.018 &  0.022 &  0.019 &  0.018 &  0.017 &  0.017 &  0.020 &  0.012 &  0.017        \\
     3.. & 17.611 & 17.552 & 17.672 & 17.589 & 17.660 & 17.727 & 17.753 & 17.836 & 17.841 & 17.868 & 17.976 & 17.923 & 17.956\\
 & 0.030 &  0.029 &  0.028 &  0.031 &  0.032 &  0.039 &  0.038 &  0.044 &  0.046 &  0.064 &  0.108 &  0.062 &  0.105        \\
     4.. & 17.879 & 17.816 & 17.853 & 17.815 & 17.824 & 17.799 & 17.782 & 17.723 & 17.864 & 17.898 & 18.029 & 17.795 & 17.970\\
 & 0.037 &  0.036 &  0.032 &  0.038 &  0.037 &  0.042 &  0.039 &  0.040 &  0.048 &  0.067 &  0.118 &  0.056 &  0.110        \\
     5.. & 18.864 & 18.685 & 18.687 & 18.415 & 18.706 & 18.537 & 18.517 & 18.612 & 18.518 & 18.800 & 19.165 & 18.925 & 18.677\\
 & 0.085 &  0.079 &  0.077 &  0.082 &  0.085 &  0.086 &  0.089 &  0.091 &  0.101 &  0.114 &  0.161 &  0.155 &  0.168        \\
     6.. & 17.611 & 17.552 & 17.672 & 17.589 & 17.660 & 17.727 & 17.753 & 17.836 & 17.841 & 17.868 & 17.976 & 17.923 & 17.956\\
 & 0.030 &  0.029 &  0.028 &  0.031 &  0.032 &  0.039 &  0.038 &  0.044 &  0.046 &  0.064 &  0.108 &  0.062 &  0.105        \\
     7.. & 16.990 & 16.874 & 16.660 & 17.116 & 17.211 & 17.338 & 16.375 & 17.337 & 17.570 & 17.668 & 17.697 & 17.187 & 18.227\\
 & 0.019 &  0.018 &  0.013 &  0.022 &  0.023 &  0.029 &  0.014 &  0.030 &  0.038 &  0.054 &  0.085 &  0.033 &  0.136        \\
     8.. & 19.042 & 18.912 & 18.910 & 18.915 & 18.896 & 18.941 & 18.179 & 18.664 & 18.759 & 18.852 & 18.688 & 18.646 & 19.728\\
 & 0.083 &  0.083 &  0.083 &  0.083 &  0.078 &  0.081 &  0.075 &  0.076 &  0.079 &  0.083 &  0.094 &  0.088 &  0.129        \\
     9.. & 20.062 & 20.574 & 20.157 & 20.084 & 19.921 & 20.203 & 20.005 & 20.255 & 19.745 & 19.620 & 20.314 & 19.683 & 19.736\\
 & 0.106 &  0.108 &  0.098 &  0.094 &  0.098 &  0.104 &  0.099 &  0.114 &  0.115 &  0.130 &  0.182 &  0.160 &  0.161        \\
    10.. & 18.649 & 18.452 & 18.515 & 18.578 & 18.522 & 18.536 & 18.618 & 18.536 & 18.603 & 18.552 & 18.400 & 18.516 & 18.179\\
 & 0.066 &  0.056 &  0.051 &  0.066 &  0.062 &  0.074 &  0.073 &  0.074 &  0.082 &  0.114 &  0.156 &  0.100 &  0.126        \\
    11.. & 19.561 & 19.051 & 18.994 & 18.823 & 18.500 & 18.520 & 18.186 & 18.114 & 17.985 & 17.948 & 17.673 & 17.618 & 17.539\\
 & 0.153 &  0.098 &  0.080 &  0.085 &  0.063 &  0.075 &  0.052 &  0.053 &  0.050 &  0.067 &  0.081 &  0.046 &  0.071        \\
    12.. & 19.485 & 19.382 & 19.468 & 19.233 & 19.302 & 19.182 & 19.115 & 19.037 & 18.911 & 18.796 & 18.902 & 18.617 & 18.305\\
 & 0.092 &  0.091 &  0.092 &  0.091 &  0.096 &  0.094 &  0.101 &  0.100 &  0.105 &  0.098 &  0.111 &  0.107 &  0.109        \\
    13.. & 19.226 & 18.983 & 19.020 & 19.092 & 18.906 & 19.128 & 19.545 & 19.113 & 19.407 & 18.900 & 19.446 & 18.868 & 19.324\\
 & 0.106 &  0.087 &  0.077 &  0.101 &  0.084 &  0.121 &  0.165 &  0.120 &  0.165 &  0.146 &  0.380 &  0.131 &  0.335        \\
    14.. & 17.411 & 17.287 & 17.291 & 17.287 & 17.252 & 17.209 & 17.275 & 17.181 & 17.089 & 17.179 & 17.087 & 17.134 & 17.088\\
 & 0.024 &  0.022 &  0.019 &  0.023 &  0.021 &  0.023 &  0.024 &  0.023 &  0.023 &  0.032 &  0.045 &  0.029 &  0.044        \\
    15.. & 19.634 & 19.213 & 19.133 & 19.111 & 18.995 & 19.099 & 19.196 & 19.203 & 18.917 & 18.928 & 18.902 & 18.788 & 19.156\\
 & 0.150 &  0.105 &  0.084 &  0.102 &  0.089 &  0.115 &  0.119 &  0.128 &  0.104 &  0.145 &  0.222 &  0.118 &  0.276        \\
    16.. & 17.302 & 17.133 & 17.190 & 17.234 & 17.218 & 17.254 & 17.211 & 17.186 & 17.209 & 17.206 & 17.242 & 17.237 & 17.204\\
 & 0.022 &  0.019 &  0.017 &  0.022 &  0.021 &  0.025 &  0.022 &  0.023 &  0.025 &  0.034 &  0.055 &  0.032 &  0.052        \\
    18.. & 19.821 & 19.538 & 19.594 & 19.689 & 19.602 & 19.673 & 19.655 & 19.631 & 19.403 & 19.543 & 19.703 & 19.433 & 18.885\\
 & 0.161 &  0.126 &  0.113 &  0.151 &  0.143 &  0.180 &  0.161 &  0.172 &  0.147 &  0.246 &  0.455 &  0.202 &  0.210        \\
\end{tabular}
\end{table}
 
\setcounter{table}{1}
\begin{table}[ht]
\caption{Continued}
\vspace {0.3cm}
\begin{tabular}{cccccccccccccc}
\hline
\hline
 No. & 03  &  04 &  05 &  06 &  07 &  08 &  09 &  10 &  11 &  12 &  13 &  14 &  15\\
(1)    & (2) & (3) & (4) & (5) & (6) & (7) & (8) & (9) & (10) & (11) & (12) & (13) & (14)\\
\hline
    19.. & 18.973 & 18.775 & 18.757 & 18.681 & 18.570 & 18.655 & 18.523 & 18.496 & 18.497 & 18.320 & 18.297 & 18.237 & 18.089\\
 & 0.088 &  0.074 &  0.063 &  0.072 &  0.065 &  0.083 &  0.068 &  0.072 &  0.076 &  0.093 &  0.143 &  0.079 &  0.116        \\
    20.. & 19.110 & 18.755 & 18.704 & 18.453 & 18.307 & 18.254 & 18.050 & 17.831 & 17.698 & 17.706 & 17.522 & 17.392 & 17.303\\
 & 0.102 &  0.077 &  0.063 &  0.063 &  0.054 &  0.061 &  0.048 &  0.043 &  0.040 &  0.055 &  0.072 &  0.038 &  0.058        \\
    21.. & 17.265 & 17.317 & 17.354 & 17.469 & 17.426 & 17.630 & 17.702 & 18.012 & 17.929 & 17.869 & 18.319 & 18.026 & 18.402\\
 & 0.023 &  0.024 &  0.021 &  0.029 &  0.025 &  0.035 &  0.036 &  0.050 &  0.049 &  0.060 &  0.139 &  0.065 &  0.149        \\
    22.. & 18.655 & 18.231 & 18.126 & 18.008 & 17.760 & 17.794 & 17.739 & 17.634 & 17.601 & 17.492 & 17.561 & 17.490 & 17.322\\
 & 0.065 &  0.046 &  0.037 &  0.040 &  0.032 &  0.038 &  0.035 &  0.034 &  0.035 &  0.043 &  0.071 &  0.040 &  0.056        \\
    23.. & 18.988 & 18.756 & 18.676 & 18.581 & 18.335 & 18.407 & 18.389 & 18.355 & 18.282 & 18.206 & 18.341 & 18.107 & 18.239\\
 & 0.085 &  0.071 &  0.057 &  0.065 &  0.051 &  0.064 &  0.059 &  0.062 &  0.062 &  0.079 &  0.139 &  0.067 &  0.125        \\
    24.. & 18.886 & 18.877 & 18.871 & 18.996 & 18.950 & 19.111 & 19.204 & 19.400 & 19.688 & 19.831 & 20.399 & 20.540 & 20.083\\
 & 0.081 &  0.082 &  0.071 &  0.097 &  0.093 &  0.127 &  0.129 &  0.166 &  0.228 &  0.372 &  0.990 &  0.654 &  0.731        \\
    25.. & 19.193 & 19.220 & 19.285 & 18.817 & 18.833 & 18.871 & 18.880 & 19.007 & 18.573 & 18.792 & 18.779 & 19.118 & 18.834\\
 & 0.079 &  0.068 &  0.073 &  0.061 &  0.071 &  0.074 &  0.079 &  0.072 &  0.070 &  0.092 &  0.188 &  0.145 &  0.173        \\
    26.. & 17.465 & 17.372 & 17.481 & 17.517 & 17.407 & 17.557 & 17.567 & 17.558 & 17.514 & 17.426 & 17.391 & 17.389 & 17.411\\
 & 0.029 &  0.028 &  0.027 &  0.035 &  0.028 &  0.038 &  0.038 &  0.040 &  0.041 &  0.045 &  0.063 &  0.041 &  0.065        \\
    27.. & 18.789 & 18.497 & 18.494 & 18.400 & 18.209 & 18.202 & 18.145 & 18.109 & 17.984 & 17.902 & 17.762 & 17.543 & 17.443\\
 & 0.088 &  0.071 &  0.061 &  0.072 &  0.055 &  0.064 &  0.059 &  0.062 &  0.059 &  0.065 &  0.083 &  0.045 &  0.063        \\
    28.. & 17.258 & 16.754 & 16.600 & 16.458 & 16.163 & 16.149 & 15.967 & 15.914 & 15.866 & 15.777 & 15.722 & 15.689 & 15.660\\
 & 0.024 &  0.017 &  0.013 &  0.014 &  0.010 &  0.012 &  0.010 &  0.010 &  0.010 &  0.011 &  0.015 &  0.010 &  0.014        \\
    29.. & 18.982 & 18.598 & 18.614 & 18.313 & 18.267 & 18.184 & 18.170 & 18.163 & 18.229 & 18.134 & 18.173 & 18.077 & 18.218\\
 & 0.056 &  0.049 &  0.043 &  0.041 &  0.036 &  0.038 &  0.039 &  0.043 &  0.052 &  0.051 &  0.055 &  0.062 &  0.088        \\
    30.. & 18.703 & 18.186 & 18.486 & 18.273 & 18.108 & 18.176 & 18.888 & 18.329 & 18.378 & 18.404 & 18.616 & 18.651 & 18.769\\
 & 0.070 &  0.046 &  0.052 &  0.053 &  0.043 &  0.053 &  0.101 &  0.062 &  0.069 &  0.093 &  0.173 &  0.108 &  0.197        \\
    31.. & 19.396 & 18.951 & 18.873 & 18.770 & 18.528 & 18.464 & 18.438 & 18.375 & 18.271 & 18.212 & 18.033 & 18.474 & 18.192\\
 & 0.129 &  0.089 &  0.071 &  0.081 &  0.064 &  0.071 &  0.065 &  0.067 &  0.065 &  0.083 &  0.109 &  0.097 &  0.125        \\
    32.. & 16.614 & 16.507 & 16.573 & 16.644 & 16.643 & 16.689 & 16.666 & 16.651 & 16.601 & 16.538 & 16.542 & 16.488 & 16.424\\
 & 0.015 &  0.014 &  0.012 &  0.016 &  0.015 &  0.017 &  0.016 &  0.017 &  0.017 &  0.021 &  0.031 &  0.018 &  0.028        \\
    33.. & 17.396 & 17.141 & 17.196 & 17.037 & 17.043 & 16.915 & 16.920 & 16.828 & 16.877 & 16.986 & 16.808 & 16.912 & 16.969\\
 & 0.024 &  0.020 &  0.018 &  0.019 &  0.018 &  0.019 &  0.018 &  0.018 &  0.020 &  0.027 &  0.035 &  0.024 &  0.040        \\
    34.. & 17.663 & 17.560 & 17.613 & 17.587 & 17.601 & 17.647 & 17.660 & 17.715 & 17.859 & 17.867 & 17.772 & 17.894 & 17.754\\
 & 0.032 &  0.029 &  0.026 &  0.031 &  0.030 &  0.036 &  0.034 &  0.038 &  0.046 &  0.063 &  0.089 &  0.059 &  0.087        \\
    35.. & 17.204 & 17.045 & 17.076 & 17.138 & 17.090 & 17.181 & 17.172 & 17.211 & 17.285 & 17.253 & 17.300 & 17.208 & 17.313\\
 & 0.019 &  0.017 &  0.015 &  0.019 &  0.018 &  0.022 &  0.020 &  0.023 &  0.025 &  0.034 &  0.055 &  0.030 &  0.055        \\
\end{tabular}
\end{table}
 
\setcounter{table}{1}
\begin{table}[ht]
\caption{Continued}
\vspace {0.3cm}
\begin{tabular}{cccccccccccccc}
\hline
\hline
 No. & 03  &  04 &  05 &  06 &  07 &  08 &  09 &  10 &  11 &  12 &  13 &  14 &
15\\
(1)    & (2) & (3) & (4) & (5) & (6) & (7) & (8) & (9) & (10) & (11) & (12) & (13) & (14)\\
\hline
    36.. & 18.771 & 18.303 & 18.563 & 18.257 & 18.468 & 18.420 & 18.363 & 18.111 & 18.163 & 18.921 & 19.179 & 18.687 & 18.599\\
 & 0.071 &  0.048 &  0.051 &  0.048 &  0.057 &  0.064 &  0.056 &  0.048 &  0.054 &  0.154 &  0.308 &  0.112 &  0.179        \\
    37.. & 18.779 & 18.613 & 19.215 & 18.490 & 18.518 & 18.541 & 18.821 & 18.276 & 18.167 & 18.190 & 18.124 & 18.023 & 17.643\\
 & 0.079 &  0.069 &  0.110 &  0.065 &  0.067 &  0.079 &  0.112 &  0.063 &  0.061 &  0.087 &  0.128 &  0.070 &  0.081        \\
    38.. & 19.003 & 18.854 & 18.737 & 18.722 & 18.742 & 18.748 & 18.641 & 18.767 & 18.783 & 18.636 & 18.545 & 18.725 & 18.201\\
 & 0.093 &  0.083 &  0.064 &  0.079 &  0.079 &  0.094 &  0.080 &  0.096 &  0.104 &  0.126 &  0.181 &  0.126 &  0.131        \\
    40.. & 17.711 & 17.656 & 17.540 & 17.919 & 18.069 & 18.182 & 17.369 & 18.177 & 18.494 & 18.309 & 18.707 & 18.382 & 18.474\\
 & 0.024 &  0.023 &  0.019 &  0.028 &  0.031 &  0.038 &  0.020 &  0.039 &  0.046 &  0.068 &  0.120 &  0.063 &  0.117        \\
    43.. & 18.437 & 18.357 & 18.236 & 18.387 & 18.424 & 18.531 & 17.687 & 18.481 & 18.450 & 18.346 & 18.555 & 17.908 & 18.032\\
 & 0.044 &  0.041 &  0.034 &  0.034 &  0.035 &  0.034 &  0.018 &  0.038 &  0.032 &  0.035 &  0.052 &  0.032 &  0.037        \\
    44.. & 17.629 & 17.556 & 17.624 & 17.642 & 17.607 & 17.658 & 17.679 & 17.669 & 17.727 & 17.688 & 17.634 & 17.576 & 17.486\\
 & 0.030 &  0.029 &  0.026 &  0.032 &  0.030 &  0.037 &  0.035 &  0.038 &  0.042 &  0.054 &  0.079 &  0.045 &  0.069        \\
    45.. & 19.634 & 19.213 & 19.133 & 19.111 & 18.995 & 19.099 & 19.196 & 19.203 & 18.917 & 18.928 & 18.902 & 18.788 & 19.156\\
 & 0.150 &  0.105 &  0.084 &  0.102 &  0.089 &  0.115 &  0.119 &  0.128 &  0.104 &  0.145 &  0.222 &  0.118 &  0.276        \\
    46.. & 19.400 & 18.828 & 18.827 & 18.727 & 18.499 & 18.617 & 18.559 & 18.487 & 18.643 & 18.419 & 18.505 & 18.379 & 18.839\\
 & 0.133 &  0.082 &  0.071 &  0.080 &  0.063 &  0.082 &  0.074 &  0.074 &  0.090 &  0.101 &  0.169 &  0.090 &  0.227        \\
    47.. & 17.547 & 17.357 & 17.414 & 17.382 & 17.214 & 17.227 & 17.183 & 17.165 & 17.209 & 17.296 & 17.288 & 17.159 & 17.220\\
 & 0.025 &  0.023 &  0.021 &  0.025 &  0.023 &  0.027 &  0.023 &  0.025 &  0.027 &  0.039 &  0.059 &  0.032 &  0.059        \\
    48.. & 18.878 & 18.853 & 18.967 & 18.862 & 18.691 & 18.790 & 19.521 & 18.639 & 18.805 & 18.790 & 18.593 & 18.692 & 18.948\\
 & 0.088 &  0.089 &  0.084 &  0.095 &  0.075 &  0.096 &  0.185 &  0.087 &  0.108 &  0.136 &  0.172 &  0.118 &  0.238        \\
    49.. & 19.374 & 18.537 & 18.556 & 18.518 & 18.081 & 18.068 & 17.862 & 17.905 & 17.928 & 17.743 & 17.797 & 17.810 & 17.752\\
 & 0.151 &  0.073 &  0.065 &  0.078 &  0.049 &  0.057 &  0.046 &  0.050 &  0.054 &  0.059 &  0.094 &  0.059 &  0.090        \\
    50.. & 18.633 & 18.272 & 18.249 & 18.411 & 18.227 & 18.282 & 18.205 & 18.457 & 18.367 & 18.246 & 18.172 & 18.094 & 18.173\\
 & 0.065 &  0.049 &  0.041 &  0.059 &  0.048 &  0.059 &  0.052 &  0.071 &  0.069 &  0.083 &  0.120 &  0.067 &  0.119        \\
    51.. & 18.626 & 18.513 & 18.554 & 18.519 & 18.446 & 18.375 & 18.508 & 18.317 & 18.475 & 18.506 & 18.553 & 19.044 & 18.452\\
 & 0.080 &  0.073 &  0.077 &  0.093 &  0.102 &  0.098 &  0.092 &  0.100 &  0.082 &  0.109 &  0.176 &  0.081 &  0.107        \\
    52.. & 19.742 & 19.233 & 19.102 & 19.202 & 18.741 & 18.901 & 18.707 & 18.744 & 18.805 & 18.592 & 18.570 & 18.688 & 18.253\\
 & 0.161 &  0.104 &  0.080 &  0.109 &  0.072 &  0.098 &  0.077 &  0.086 &  0.096 &  0.111 &  0.170 &  0.111 &  0.126        \\
    53.. & 20.342 & 19.691 & 19.543 & 19.240 & 19.240 & 19.054 & 18.938 & 18.800 & 18.634 & 18.488 & 18.661 & 18.481 & 18.162\\
 & 0.285 &  0.157 &  0.118 &  0.108 &  0.115 &  0.114 &  0.092 &  0.088 &  0.080 &  0.108 &  0.204 &  0.097 &  0.126        \\
    54.. & 19.080 & 18.806 & 18.628 & 18.401 & 18.217 & 18.071 & 17.889 & 17.803 & 17.820 & 17.716 & 17.730 & 17.586 & 17.648\\
 & 0.088 &  0.069 &  0.051 &  0.050 &  0.045 &  0.046 &  0.036 &  0.036 &  0.038 &  0.052 &  0.084 &  0.042 &  0.076        \\
    55.. & 18.598 & 18.248 & 18.212 & 18.174 & 18.005 & 18.024 & 18.027 & 17.939 & 17.940 & 17.884 & 17.797 & 17.659 & 17.742\\
 & 0.060 &  0.045 &  0.038 &  0.045 &  0.038 &  0.045 &  0.043 &  0.042 &  0.045 &  0.059 &  0.084 &  0.044 &  0.079        \\
\end{tabular}
\end{table}
 
\setcounter{table}{1}
\begin{table}[ht]
\caption{Continued}
\vspace {0.3cm}
\begin{tabular}{cccccccccccccc}
\hline
\hline
 No. & 03  &  04 &  05 &  06 &  07 &  08 &  09 &  10 &  11 &  12 &  13 &  14 &
15\\
(1)    & (2) & (3) & (4) & (5) & (6) & (7) & (8) & (9) & (10) & (11) & (12) & (13) & (14)\\
\hline
    56.. & 18.662 & 18.454 & 18.469 & 18.414 & 18.308 & 18.342 & 18.362 & 18.343 & 18.471 & 18.437 & 18.508 & 18.196 & 18.627\\
 & 0.068 &  0.058 &  0.051 &  0.060 &  0.054 &  0.065 &  0.061 &  0.065 &  0.078 &  0.105 &  0.175 &  0.078 &  0.193        \\
    57.. & 19.939 & 19.766 & 19.506 & 19.471 & 18.933 & 19.041 & 18.786 & 18.848 & 18.532 & 18.321 & 18.256 & 18.157 & 18.248\\
 & 0.206 &  0.177 &  0.120 &  0.144 &  0.091 &  0.118 &  0.086 &  0.099 &  0.080 &  0.095 &  0.142 &  0.075 &  0.139        \\
    58.. & 18.762 & 18.537 & 18.585 & 18.546 & 18.387 & 18.478 & 18.388 & 18.534 & 18.572 & 18.430 & 18.127 & 18.359 & 18.628\\
 & 0.074 &  0.061 &  0.055 &  0.065 &  0.057 &  0.072 &  0.061 &  0.076 &  0.083 &  0.105 &  0.126 &  0.090 &  0.197        \\
    59.. & 18.467 & 18.294 & 18.284 & 18.194 & 17.923 & 17.976 & 17.872 & 17.689 & 17.484 & 17.382 & 17.338 & 17.155 & 17.045\\
 & 0.046 &  0.038 &  0.034 &  0.037 &  0.033 &  0.036 &  0.031 &  0.029 &  0.026 &  0.034 &  0.050 &  0.028 &  0.041        \\
    60.. & 18.798 & 18.714 & 18.664 & 18.648 & 18.676 & 18.703 & 18.689 & 18.783 & 18.806 & 18.713 & 19.077 & 18.649 & 18.950\\
 & 0.074 &  0.070 &  0.057 &  0.068 &  0.071 &  0.085 &  0.077 &  0.091 &  0.099 &  0.131 &  0.291 &  0.113 &  0.255        \\
\hline
\end{tabular}
\end{table}
} 


\setcounter{table}{2}
\begin{table}[ht]
\caption{Comparison of Cluster Photometry with Previous Measurements}
\vspace {0.3cm}
\begin{tabular}{ccccc}
\hline
\hline
No. & $V$ (Chandar et al.)  & $V$ (BATC) &$B-V$ (Chandar et al.) &$B-V$ (BATC) \\
(1)    & (2) & (3) & (4) & (5) \\
\hline
      1...... & 18.594 $\pm$  0.018 & 18.722 $\pm$  0.148 & 0.028 $\pm$  0.020 & -0.005 $\pm$  0.275\\
      2...... & 16.756 $\pm$  0.006 & 16.696 $\pm$  0.032 & 0.214 $\pm$  0.006 & -0.015 $\pm$  0.060\\
      3...... & 17.416 $\pm$  0.010 & 17.674 $\pm$  0.054 & 0.294 $\pm$  0.023 & -0.062 $\pm$  0.097\\
      4...... & 17.851 $\pm$  0.006 & 17.888 $\pm$  0.063 & 0.176 $\pm$  0.011 & 0.008 $\pm$  0.114\\
      5...... & 18.621 $\pm$  0.013 & 18.726 $\pm$  0.139 & 0.068 $\pm$  0.019 & 0.073 $\pm$  0.254\\
      6...... & 17.862 $\pm$  0.008 & 17.674 $\pm$  0.054 & -0.186 $\pm$  0.013 & -0.062 $\pm$  0.097\\
      7...... & 17.384 $\pm$  0.006 & 17.198 $\pm$  0.039 & -0.131 $\pm$  0.006 & -0.224 $\pm$  0.069\\
      8...... & 18.856 $\pm$  0.019 & 18.947 $\pm$  0.131 & 0.352 $\pm$  0.030 & 0.069 $\pm$  0.250\\
      9...... & 19.670 $\pm$  0.032 & 19.942 $\pm$  0.161 &                   & 0.686 $\pm$  0.314\\
     10...... & 18.774 $\pm$  0.016 & 18.595 $\pm$  0.108 &                   & -0.039 $\pm$  0.190\\
     11...... & 18.808 $\pm$  0.016 & 18.657 $\pm$  0.115 &                   & 0.594 $\pm$  0.265\\
     12...... & 19.008 $\pm$  0.021 & 19.377 $\pm$  0.155 & 0.297 $\pm$  0.034 & 0.082 $\pm$  0.288\\
     13...... & 18.667 $\pm$  0.019 & 18.953 $\pm$  0.156 & 0.208 $\pm$  0.016 & 0.149 $\pm$  0.283\\
     14...... & 17.339 $\pm$  0.005 & 17.336 $\pm$  0.036 & 0.170 $\pm$  0.008 & 0.051 $\pm$  0.067\\
     15...... & 19.008 $\pm$  0.023 & 19.058 $\pm$  0.159 & 0.409 $\pm$  0.028 & 0.340 $\pm$  0.316\\
     16...... & 17.449 $\pm$  0.005 & 17.271 $\pm$  0.036 & 0.267 $\pm$  0.009 & -0.039 $\pm$  0.064\\
     18...... & 19.870 $\pm$  0.046 & 19.666 $\pm$  0.250 & 0.007 $\pm$  0.038 & -0.004 $\pm$  0.437\\
     19...... & 18.752 $\pm$  0.016 & 18.637 $\pm$  0.116 &                   & 0.260 $\pm$  0.223\\
     20...... & 18.302 $\pm$  0.005 & 18.430 $\pm$  0.094 &                   & 0.489 $\pm$  0.208\\
     21...... & 17.713 $\pm$  0.008 & 17.433 $\pm$  0.046 & -0.190 $\pm$  0.008 & -0.062 $\pm$  0.080\\
     22...... & 17.935 $\pm$  0.007 & 17.888 $\pm$  0.058 & 0.513 $\pm$  0.016 & 0.534 $\pm$  0.126\\
     23...... & 18.264 $\pm$  0.010 & 18.450 $\pm$  0.093 & 0.548 $\pm$  0.017 & 0.449 $\pm$  0.195\\
     24...... & 18.904 $\pm$  0.017 & 18.972 $\pm$  0.166 & -0.044 $\pm$  0.016 & -0.017 $\pm$  0.281\\
     25...... & 18.840 $\pm$  0.018 & 18.875 $\pm$  0.114 & 0.692 $\pm$  0.039 & 0.399 $\pm$  0.216\\
     26...... & 17.535 $\pm$  0.022 & 17.453 $\pm$  0.052 & 0.172 $\pm$  0.027 & -0.011 $\pm$  0.093\\
     27...... & 18.077 $\pm$  0.022 & 18.332 $\pm$  0.098 & 0.273 $\pm$  0.035 & 0.305 $\pm$  0.203\\
     28...... & 16.368 $\pm$  0.015 & 16.322 $\pm$  0.019 & 0.794 $\pm$  0.014 & 0.652 $\pm$  0.044\\
     29...... & 18.437 $\pm$  0.035 & 18.368 $\pm$  0.062 & 0.985 $\pm$  0.059 & 0.386 $\pm$  0.133\\
     30...... & 18.124 $\pm$  0.041 & 18.198 $\pm$  0.077 & 0.273 $\pm$  0.037 & 0.110 $\pm$  0.150\\
     31...... & 18.508 $\pm$  0.031 & 18.686 $\pm$  0.113 & 0.428 $\pm$  0.040 & 0.455 $\pm$  0.247\\
     32...... & 16.703 $\pm$  0.016 & 16.687 $\pm$  0.025 & 0.079 $\pm$  0.013 & -0.097 $\pm$  0.045\\
\end{tabular}
\end{table} 
 
\setcounter{table}{2}
\begin{table}[ht]
\caption{Continued}
\vspace {0.3cm}
\begin{tabular}{ccccc}
\hline
\hline
No. & $V$ (Chandar et al.)  & $V$ (BATC) &$B-V$ (Chandar et al.) &$B-V$ (BATC)\\
(1)    & (2) & (3) & (4) & (5) \\
\hline
     33...... & 17.141 $\pm$  0.020 & 17.141 $\pm$  0.031 & 0.218 $\pm$  0.018 & 0.118 $\pm$  0.060\\
     34...... & 17.722 $\pm$  0.021 & 17.641 $\pm$  0.051 & 0.137 $\pm$  0.020 & 0.005 $\pm$  0.094\\
     35...... & 17.145 $\pm$  0.017 & 17.135 $\pm$  0.031 & 0.150 $\pm$  0.015 & 0.012 $\pm$  0.056\\
     36...... & 18.553 $\pm$  0.034 & 18.474 $\pm$  0.093 & 0.222 $\pm$  0.033 & -0.051 $\pm$  0.168\\
     37...... & 18.636 $\pm$  0.033 & 18.561 $\pm$  0.113 & 0.244 $\pm$  0.033 & 0.163 $\pm$  0.230\\
     38...... & 18.655 $\pm$  0.032 & 18.793 $\pm$  0.135 & 0.202 $\pm$  0.040 & 0.195 $\pm$  0.253\\
     40...... & 18.505 $\pm$  0.010 & 18.043 $\pm$  0.052 & -0.338 $\pm$  0.010 & -0.301 $\pm$  0.125\\
     43...... & 18.575 $\pm$  0.018 & 18.436 $\pm$  0.057 & 0.755 $\pm$  0.039 & 0.012 $\pm$  0.115\\
     44...... & 17.735 $\pm$  0.007 & 17.661 $\pm$  0.052 & 0.288 $\pm$  0.009 & -0.030 $\pm$  0.093\\
     45...... & 19.074 $\pm$  0.024 & 19.058 $\pm$  0.159 & 0.535 $\pm$  0.026 & 0.340 $\pm$  0.316\\
     46...... & 18.474 $\pm$  0.013 & 18.594 $\pm$  0.116 & 0.764 $\pm$  0.017 & 0.436 $\pm$  0.243\\
     47...... & 17.264 $\pm$  0.005 & 17.323 $\pm$  0.040 & 0.497 $\pm$  0.007 & 0.137 $\pm$  0.072\\
     48...... & 18.794 $\pm$  0.100 & 18.773 $\pm$  0.137 & 0.417 $\pm$  0.026 & 0.134 $\pm$  0.264\\
     49...... & 18.079 $\pm$  0.011 & 18.285 $\pm$  0.092 & 0.824 $\pm$  0.017 & 0.507 $\pm$  0.213\\
     50...... & 18.132 $\pm$  0.006 & 18.328 $\pm$  0.087 & 0.360 $\pm$  0.014 & 0.104 $\pm$  0.159\\
     51...... & 18.838 $\pm$  0.022 & 18.552 $\pm$  0.164 & 0.459 $\pm$  0.026 & 0.052 $\pm$  0.272\\
     52...... & 18.618 $\pm$  0.016 & 18.897 $\pm$  0.139 & 0.810 $\pm$  0.027 & 0.552 $\pm$  0.296\\
     53...... & 19.563 $\pm$  0.027 & 19.359 $\pm$  0.187 &                   & 0.583 $\pm$  0.434\\
     54...... & 18.407 $\pm$  0.010 & 18.383 $\pm$  0.076 & 1.075 $\pm$  0.020 & 0.598 $\pm$  0.176\\
     55...... & 18.104 $\pm$  0.006 & 18.112 $\pm$  0.068 &                   & 0.295 $\pm$  0.135\\
     56...... & 18.408 $\pm$  0.008 & 18.390 $\pm$  0.094 &                   & 0.181 $\pm$  0.178\\
     57...... & 19.375 $\pm$  0.015 & 19.131 $\pm$  0.176 &                   & 0.805 $\pm$  0.425\\
     58...... & 18.391 $\pm$  0.006 & 18.468 $\pm$  0.101 &                   & 0.182 $\pm$  0.190\\
     59...... & 18.649 $\pm$  0.009 & 18.052 $\pm$  0.056 &                   & 0.356 $\pm$  0.112\\
     60...... & 18.679 $\pm$  0.008 & 18.717 $\pm$  0.121 &                   & 0.101 $\pm$  0.220\\
\hline
\end{tabular}
\end{table}


\begin{table}[ht]
\caption[]{Age and Metallicity Distribution of 56 Star Clusters}
\vspace {0.3cm}
\begin{tabular}{ccc|ccc}
\hline
\hline
 No. & Metallicity ($Z$)& Age ([$\log$ yr]) & No. & Metallicity ($Z$)& Age ([$\log$ yr])\\
 (1) & (2) & (3) & (1) & (2) & (3) \\
\hline
      1...... & 0.02000 &  7.260 &     30...... & 0.00400 &  6.620\\
      2...... & 0.05000 &  6.860 &     31...... & 0.00040 &  9.207\\
      3...... & 0.00400 &  6.620 &     32...... & 0.02000 &  6.880\\
      4...... & 0.00040 &  8.009 &     33...... & 0.00400 &  8.009\\
      5...... & 0.00040 &  8.009 &     34...... & 0.00040 &  7.806\\
      6...... & 0.00400 &  6.820 &     35...... & 0.00040 &  7.806\\
      7...... & 0.00400 &  6.439 &     36...... & 0.00040 &  7.806\\
      8...... & 0.00040 &  6.620 &     37...... & 0.02000 &  6.940\\
      9...... & 0.02000 &  7.477 &     38...... & 0.02000 &  8.255\\
     10...... & 0.00400 &  7.320 &     40...... & 0.00400 &  6.420\\
     11...... & 0.00400 & 10.301 &     43...... & 0.02000 &  7.121\\
     12...... & 0.02000 &  6.940 &     44...... & 0.00400 &  7.220\\
     13...... & 0.00040 &  7.806 &     45...... & 0.00040 &  8.009\\
     14...... & 0.00400 &  6.980 &     46...... & 0.00040 &  8.957\\
     15...... & 0.00400 &  8.009 &     47...... & 0.02000 &  8.057\\
     16...... & 0.00400 &  7.079 &     48...... & 0.00040 &  8.009\\
     18...... & 0.02000 &  7.279 &     49...... & 0.00040 &  9.342\\
     19...... & 0.00400 &  6.940 &     50...... & 0.02000 &  8.057\\
     20...... & 0.02000 &  9.544 &     51...... & 0.00040 &  7.806\\
     21...... & 0.02000 &  6.580 &     52...... & 0.00040 &  9.107\\
     22...... & 0.00040 &  9.255 &     53...... & 0.00400 &  8.957\\
     23...... & 0.00040 &  8.957 &     54...... & 0.00040 &  9.009\\
     24...... & 0.02000 &  6.520 &     55...... & 0.00040 &  8.957\\
     25...... & 0.00400 &  8.009 &     56...... & 0.00040 &  8.009\\
     26...... & 0.00400 &  7.241 &     57...... & 0.00400 & 10.279\\
     27...... & 0.02000 &  8.857 &     58...... & 0.00400 &  6.960\\
     28...... & 0.00040 &  9.796 &     59...... & 0.02000 &  9.107\\
     29...... & 0.00400 &  8.009 &     60...... & 0.00040 &  7.806\\
\hline
\end{tabular}
\end{table}

\begin{table}[ht]
\caption[]{Metallicity, Reddening and Age Distribution of 7 Star Clusters}
\vspace {0.3cm}
\begin{tabular}{cccc}
\hline
\hline
 No. & Metallicity ($Z$) & $E(B-V)$ & Age ([$\log$ yr])\\
\hline
   11...... & 0.00099 & 0.200 & 10.30 \\
   20...... & 0.00719 & 0.110 & 10.30 \\
   22...... & 0.00186 & 0.140 & 9.23 \\
   28...... & 0.00023 & 0.050 & 10.30 \\
   49...... & 0.00040 & 0.010 & 9.92 \\
   54...... & 0.00890 & 0.010 & 9.40 \\
   59...... & 0.01447 & 0.010 & 10.28 \\
\hline
\end{tabular}
\end{table}

\begin{thebibliography}{}

\bibitem[Allard \& Hauschildt 1995]{Allard95} 
Allard, F., \& Hauschildt, P. H. 1995, \apj, 445, 433

\bibitem[Bessell et al. 1991]{Bessell91}
Bessell, M. S., Brett, J. M., Scholz, M., \& Wood, P. R. 1991, \aaps, 89, 335

\bibitem{}Bertelli, G., Bressan, A., Chiosi, C., Fagotto, F., \& Nasi, E.
1994, A\&AS, 106, 275

\bibitem{}Bica, E., \& Alloin, D. 1986a, A\&A, 162, 21

\bibitem{}Bica, E., \& Alloin, D. 1986b, A\&AS, 66, 171

\bibitem{}Bica, E., \& Alloin, D. 1987, A\&A, 186, 49


\bibitem[Bressan et al. 1993]{Bressan93}
Bressan, A., et al. 1993, \aaps, 100, 647

\bibitem[Bressan et al. 1996]{Bressan96} 
Bressan, A., Chiosi, C., \& Tantalo, R. 1996, \aap, 311, 425

\bibitem[Bruzual \& Charlot 1993]{Bruzual93} 
Bruzual, G., \& Charlot, S. 1993, \apj, 405, 538

\bibitem[Bruzual \& Charlot 1996]{Bruzual96} 
Bruzual, G., \& Charlot, S. 1996, Documentation for GISSEL96 Spectral Synthesis Code.

\bibitem[Buzzoni 1997]{Buzzoni97} 
Buzzoni, A. 1997, in IAU Symp. 183, Cosmological Parameters and Evolution of
the Universe, ed. K. Sato, 18
 
\bibitem[Calzetti 1997]{Calzetti97} 
Calzetti, D. 1997, \aj, 113, 162

\bibitem{}Chandar, R., Bianchi, L., \& Ford, H. C. 1999a, ApJS,
122, 431

\bibitem{}Chandar, R., Bianchi, L., \& Ford, H. C. 1999b, ApJ,
517, 668

\bibitem{}Chandar, R., Bianchi, L., Ford, H. C., \&
Salasnich, B. 1999c, PASP, 111, 794

\bibitem[Charlot \& Bruzual 1991]{Charlot91} 
Charlot, S., \& Bruzual, G. 1991, \apj, 367, 126

\bibitem[Charlot et al. 1996]{Charlot96} 
Charlot, S., Worthey, G., \& Bressan, A. 1996, \apj, 457, 625

\bibitem[Chiosi et al. 1998]{Chiosi98} 
Chiosi, C., Bressan, A., Portinari, L., \& Tantalo, R. 1998, \aap, 339, 355

\bibitem{}Christian, C. A., \& Schommer, R. A. 1982, ApJS, 49, 405

\bibitem{}Christian, C. A., \& Schommer, R. A. 1983, ApJ, 275, 92

\bibitem{}Christian, C. A., \& Schommer, R. A. 1988, AJ, 95, 704

\bibitem{}Ciani, A., D'Odorico, S., \& Benvenuti, P. 1984, A\&A, 137, 223

\bibitem[Fagotto et al. 1994]{Fagotto94} 
Fagotto, F., Bressan, A., Bertelli, G., \& Chiosi, C. 1994, \aaps, 105, 39

\bibitem[Fan et al. 1996]{Fan96} 
Fan, X.-H., et al. 1996, \aj, 112, 628

\bibitem[Fioc \& Rocca-Volmerange 1997]{Fioc97} 
Fioc, M., \& Rocca-Volmerange, B. 1997, \aap, 326,950

\bibitem[Fluks et al. 1994]{Fluks94} 
Fluks, M. A., Plez, B., Th\'{e}, P. S., de Winter, D.,
Westerlund, B. E., \& Steenman, H. C. 1994, \aaps, 105, 311

\bibitem{}Freedman, W. L., Wilson, C. D., \& Madore, B. F.
1991, ApJ, 372, 455

\bibitem{GE2000} Galad\'\i-Enr\'\i quez, D., Trullols, E., \& Jordi, C.
2000, A\&AS, 146, 169

\bibitem[Girardiet al. 1996]{Girardiet96} 
Girardi, L., Bressan, A., Chiosi, C., Bertelli, G., \&
Nasi, E. 1996,\aaps,117,113

\bibitem{}Hiltner, W. A. 1960, ApJ, 131, 163

\bibitem[Iglesias et al. 1992]{Iglesias92} 
Iglesias, C. A., Rogers, F. J., \& Wilson, B. G. 1992, \apj, 397, 717

\bibitem[Jablonka et al. 1996]{Jab96} 
Jablonka, P., Bica, E., Pelat, D., \& Alloin, D. 1996, \aap, 307, 385

\bibitem[Kennicutt 1998]{Kennicutt98}
Kennicutt, R. C. 1998, \aapr, 36, 189

\bibitem{}Kong, X. et al. 2000, AJ, 119, 2745

\bibitem{}Kron, G. E., \& Mayall, N. U. 1960, AJ, 65, 581

\bibitem[Kurucz 1995]{Kurucz95} 
Kurucz, R. L. 1995 (private communication)

\bibitem{}Landolt, A. U. 1983, AJ, 88, 439
\bibitem{}Landolt, A. U. 1992, AJ, 104, 340
\bibitem[Leitherer et al. 1996]{Leitherer96} 
Leitherer, C., et al. 1996, \pasp, 108, 996

\bibitem[Leitherer et al. 1999]{Leitherer99} 
Leitherer, C., et al. 1999, \apjs, 123, 3

\bibitem[Lejeune et al. 1997]{Lejeune97} 
Lejeune, Th., Cuisinier, F., \& Buser, R. 1997, \aaps, 125, 229

\bibitem[Lejeune et al. 1998]{Lejeune98} 
Lejeune, Th., Cuisinier, F., \& Buser, R. 1998, \aaps, 130, 65

\bibitem{}McClure, R. D., \& Racine, R. 1969, AJ, 74, 1000

\bibitem{}Melnick, J., \& D'Odorico, S. 1978, A\&AS, 34, 249
\bibitem[Moll\`{a} et al. 1997]{Molla97} 
Moll\`{a}, M., Ferrini, F., \& Diaz, A. I. 1997, \apj, 475, 519

\bibitem[Origlia et al. 1999]{Origlia99} 
Origlia, L., Goldader, J. D., Leitherer, C., \&
Schaerer, D., \& Oliva, E. 1999, \apj, 514, 96

\bibitem[Salpeter 1995]{Salpeter95} 
Salpeter, E. E. 1955, \apj, 121, 161

\bibitem{}Sarajedini, A. A., Geisler, D., Harding, P., \&
Schommer, R. 1998, ApJ, 508, L37

\bibitem[Sawicki \& Yee 1998]{Sawicki98} 
Sawicki, M., \& Yee, H. K. C. 1998, \aj, 115, 1329

\bibitem[Schaerer \& de Koter 1997]{Schaerer97} 
Schaerer, D., \& de Koter, A. 1997, \aap, 322, 598

\bibitem[Schaerer \& Vacca 1998]{Schaerer98} 
Schaerer, D., \& Vacca, W. D. 1998, \apj, 497, 618

\bibitem{}Schmidt, A. A., Bica, E., \& Alloin, D. 1990, MNRAS, 243, 620

\bibitem[Searle et al. 1973]{Searle73} 
Searle, L., Sargent, W. L. W., \& Bagnuolo, W. G. 1973, \apj, 179, 427

\bibitem[Tinsley 1972]{Tinsley71} 
Tinsley, B. M. 1972, \aap, 20, 382

\bibitem[Vazdekis et al. 1997]{Vazdekis97} 
Vazdekis, A., Peletier, R. F., Beckman, J. E., \& Casuso, E. 1997, \apjs, 111, 203

\bibitem[Zheng et al. 1999]{Zheng99} 
Zheng, Z. Y., et al. 1999, \aj, 117, 2757

\bibitem[Zhou 2001]{Zhou01}
Zhou, X., et al. 2001, in preparation

\bibitem[Zombeck 1990]{Zombeck90} 
Zombeck, M. V. 1990, Handbook of Space Astronomy and Astrophysics (2nd. ed; 
Cambrigde: Cambrigde Univ. Press) p. 104
\end{thebibliography}
\end{document}